\documentclass[lettersize,journal]{IEEEtran}
\IEEEoverridecommandlockouts
\usepackage{amsmath,amsfonts}
\usepackage{amssymb}
\usepackage[ruled]{algorithm2e}
\usepackage{array}
\usepackage[caption=false,font=normalsize,labelfont=sf,textfont=sf]{subfig}
\usepackage{textcomp}
\usepackage{stfloats}
\usepackage{url}
\usepackage{verbatim}
\usepackage{graphicx}
\usepackage{cite}
\usepackage{mathrsfs}
\usepackage{bbding}
\usepackage{fontawesome}
\usepackage{pifont}

\begin{document}

\title{\huge Holographic-Pattern Based Multi-User Beam Training in RHS-Aided Hybrid Near-Field and Far-Field Communications}

\author{
\IEEEauthorblockN
{
Shupei Zhang, \IEEEmembership{Student Member, IEEE},
Boya Di, \IEEEmembership{Member, IEEE},\\
Aryan Kaushik, \IEEEmembership{Member, IEEE},
and Yonina C. Eldar, \IEEEmembership{Fellow, IEEE}
}
\thanks{Shupei Zhang and Boya Di are with State Key Laboratory of Advanced Optical Communication Systems and Networks, School of Electronics, Peking University, Beijing, China. (emails: zhangshupei@pku.edu.cn; diboya@pku.edu.cn).}
\thanks{Aryan Kaushik is with the Department of Computing \& Mathematics, Manchester Metropolitan University, UK (email: a.kaushik@ieee.org).}
\thanks{Yonina C. Eldar is with the Faculty of Mathematics and Computer Science, Weizmann Institute of Science, Rehovot 7610001, Israel (e-mail: yonina.eldar@weizmann.ac.il).}
}

\maketitle

\begin{abstract}
Reconfigurable holographic surfaces~(RHSs) have been suggested as an energy-efficient solution for extremely large-scale arrays. By controlling the amplitude of RHS elements, high-gain directional holographic patterns can be achieved. However, the complexity of acquiring real-time channel state information (CSI) for beamforming is exceedingly high, particularly in large-scale RHS-assisted communications, where users may distribute in the near-field region of RHS. This paper proposes a one-shot multi-user beam training scheme in large-scale RHS-assisted systems applicable to both near and far fields. The proposed beam training scheme comprises two phases: angle search and distance search, both conducted simultaneously for all users.
For the angle search, an RHS angular codebook is designed based on holographic principles so that each codeword covers multiple angles in both near-field and far-field regions, enabling simultaneous angular search for all users. For the distance search, we construct the distance-adaptive codewords covering all candidate angles of users in a real-time way by leveraging the additivity of holographic patterns, which is different from the traditional phase array case.  
Simulation results demonstrate that the proposed scheme achieves higher system throughput compared to traditional beam training schemes. The beam training accuracy approaches the upper bound of exhaustive search at a significantly reduced overhead.
\end{abstract}


\section{Introduction}

Driven by the exponential growth of advanced applications, the 6G network is anticipated to support a peak data rate of 100 Gb/s,  and large-scale arrays are promising to meet such a demand~\cite{6G}.
However, traditional phased arrays (PAs) rely on high-resolution phase shifters, which incur unaffordable hardware costs and power consumption when transitioning to a large scale~\cite{XL},\cite{HDMA}.
The emergence of reconfigurable holographic surfaces~(RHSs) presents a potential breakthrough, which comprises a massive number of low-cost meta-material elements serving as transceiver antennas~\cite{Holo1}.
As a type of leaky-wave antenna, electromagnetic~(EM) waves continuously propagate along the RHS to generate \emph{holographic patterns}, enabling high directional gain.
Different from PAs where each antenna necessitates a feed line and phase-shifting components, the amplitude of each RHS element is controlled through a simple diode-based circuit, offering an energy-efficient solution~\cite{Holo2}.
Distinct from reflective metamaterial antennas known as reconfigurable intelligent surfaces (RISs), RHSs integrate the feed network within the metasurface, facilitating ultra-thin integration into transceivers~\cite{RIS1},~\cite{RIS2}.

While RHSs offer a potential solution for deploying large-scale arrays, the number of channel links also increases proportionally, which leads to significant complexity in traditional pilot-based channel estimation schemes~\cite{est1}.
To circumvent the need for accurate channel state information~(CSI), low-overhead beam training has emerged as a practical approach for beamforming~\cite{codebook1},\cite{codebook2}. Additionally, as the aperture of large-scale RHSs expands, the corresponding near-field region enlarges as well. Users may randomly distribute in both the near-field and far-field of the RHS, giving rise to \emph{hybrid near-far-field communications}.

Existing low-overhead beam training schemes primarily fall into two categories: angle-domain beam training based on plane waves~\cite{far_multi-user}\,-\!\!\cite{DFT2} and angle-distance domain beam training based on spherical waves~\cite{exh}\,-\!\!\cite{2D}. The authors in \cite{far_multi-user} design an adaptive hierarchical codebook in the angle domain, supporting low-overhead multi-user simultaneous beam training. A near-field codebook in the polar domain with distance and angle sampling criteria is proposed in \cite{exh}.
In \cite{twostage1} and \cite{dft-dis}, a two-stage near-field beam training scheme is introduced for the single-user system, which first searches the angle domain and then the distance domain at candidate angles. A hierarchical angle-distance beam training for a single user is designed in \cite{2D} by generating codewords with varying resolutions.

\renewcommand\arraystretch{1.2} 
\begin{table}[t]
\label{intro}
\centering
\caption{Comparison with existing beam training schemes.}
\resizebox{\columnwidth}{!}{%
\begin{tabular}{|c|c|c|c|c|}
\hline
\textbf{Beam training scheme}   & \textbf{Single-user}   & \textbf{Multi-user}          & \begin{tabular}[c]{@{}c@{}}\textbf{One-shot}\\ \textbf{angular search}\end{tabular}  & \begin{tabular}[c]{@{}c@{}}\textbf{One-shot}\\ \textbf{distance search}\end{tabular}\\ \hline

Near-field schemes in\cite{twostage1},\cite{twostage2},\cite{twostage3}   &\checkmark       & \ding{55} &    \ding{55} &\ding{55}     \\ \hline

Near-field schemes in\cite{multiuser1}     &\checkmark  & \checkmark        & \ding{55}  &\checkmark     \\ \hline
Far-field scheme in\cite{far_multi-user}      &  \checkmark        &  \checkmark   & \checkmark  & \ding{55} \\ \hline

Proposed hybrid-field scheme      &  \checkmark        &  \checkmark   & \checkmark  & \checkmark \\ \hline
\end{tabular}%
}
\end{table}

However, existing low-overhead near-field beam training schemes are limited to serving a single user~\cite{twostage1},\cite{twostage2},\cite{twostage3}.
Traditional far-field multi-user hierarchical training schemes achieve simultaneous beam training but consider only the angle domain, unable to provide accurate beam training for near-field users~\cite{far_multi-user}. Some initial works have considered near-field multi-user beam training \cite{multiuser1},\cite{multiuser2}, but the training overhead still scales with the number of users~\cite{multiuser1}, or the systems rely on neural networks requiring real-time updates, incurring high complexity~\cite{multiuser2}.
One-shot multi-user beam training in the angle-distance domain remains an unresolved challenge, which entails conducting only one-time low-cost search simultaneously for all users.
The existing beam training schemes are summarized in Table~\ref{intro}.

Therefore, it is necessary to design a low-overhead beam training mechanism that can be performed in a one-shot manner for all users, catering to both near-field and far-field users. The pivotal idea behind such a multi-user, near-far-field, one-shot beam training mechanism lies in the real-time and fast generation of multi-beam codewords, which can cover all the users based on their feedback during beam training. Owing to the superposition principle of holographic patterns, the RHS is well-suited for this scheme. Specifically, multiple single-beam holographic patterns of the RHS are directly superimposed to create a multi-beam holographic pattern.

In this paper, we propose a multi-user beam training scheme for near-far-field communications, leveraging holographic patterns. The core idea encompasses three aspects. \emph{First}, to reduce the training overhead, a hierarchical beam training mechanism is designed by employing holographic patterns with progressively narrower coverage areas. \emph{Second}, to support all the users in one shot, we generate multi-beam holographic patterns in real-time through superposition, based on user feedback during beam training process. \emph{Third}, to serve both near-field and far-field users simultaneously, the RHS codebook design takes into account the propagation characteristics of EM waves in the hybrid fields. The contributions of this paper are summarized as follows:
\begin{itemize}
\item[1)] Based on holographic principles, we propose a generalized design method for RHS multi-beam codewords. Utilizing this scheme, the RHS multi-user angular codebook is presented, which can support a one-shot angle search for both near-field and far-field users.
\item[2)] Utilizing the holographic superposition principle, we devise an RHS distance-adaptive multi-beam codebook, where each codeword covers a specific distance across all user angles.
These multi-beam codewords are generated through direct superposition of holographic patterns from single-beam codewords, enabling real-time multi-user distance search during beam training.
\item[3)] Employing the angular codebook and distance-adaptive codebook, a multi-user beam training scheme for the near-far fields is introduced.
This scheme consists of angle search and distance search phases conducted simultaneously for all users in each phase. The overhead does not increase with the number of users.
\item[4)] Simulation results show that the multi-beam codewords in both codebooks exhibit coverage aligned with the design objectives.
The proposed scheme achieves a performance close to the upper bound of exhaustive search at a significantly reduced overhead.
A higher throughput can be achieved compared to traditional near-field beam training schemes.
Moreover, this scheme is applicable to traditional far/near-field scenarios.
\end{itemize}

\emph{Organization:} The rest of this paper is organized as follows.
In Section~\ref{Sec:model}, the RHS-aided near-far-field communication is described, followed by the RHS model, the signal model, and the problem formulation.
The framework of the RHS multi-user codebook and beam training is present in Section~\ref{Sec:framework}.
The design of multi-user angular-domain codebook and distance-domain codebook is introduced in Section~\ref{Sec:angular_book} and Section~\ref{Sec:distance_book}, respectively.
In Section~\ref{Sec:beamtraining}, an RHS-aided simultaneous multi-user beam training scheme is proposed.
The simulation results are presented in Section~\ref{Sec:simulation}, and the conclusions are drawn in Section~\ref{Sec:conclusions}.

\begin{figure}[t]
\setlength{\abovecaptionskip}{0pt}
\setlength{\belowcaptionskip}{0pt}
	\centering
    \includegraphics[width=0.48\textwidth]{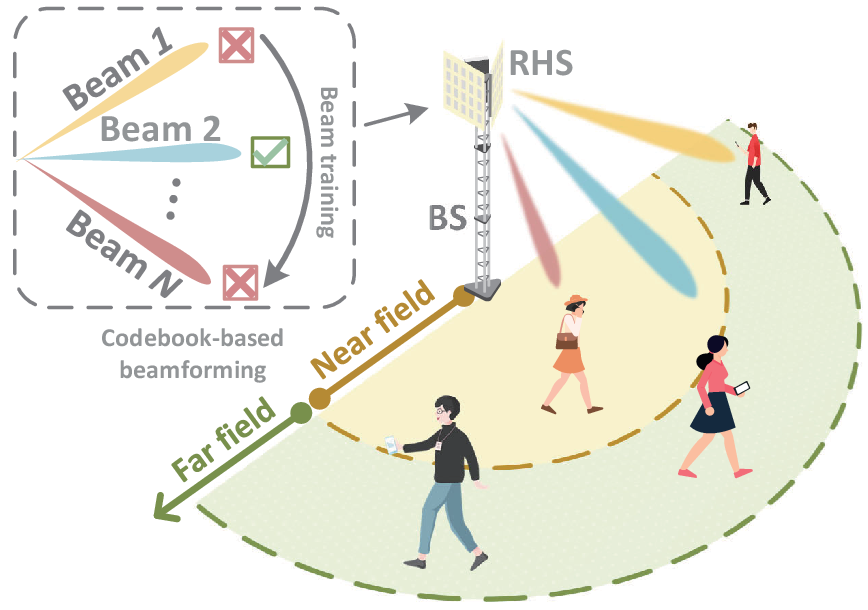}
	\caption{RHS-aided near-far-field communications.}
	\label{Fig:scenario}
\vspace{0em}
\end{figure}

\begin{figure}[t]
\setlength{\abovecaptionskip}{0pt}
\setlength{\belowcaptionskip}{0pt}
	\centering
    \includegraphics[width=0.5\textwidth]{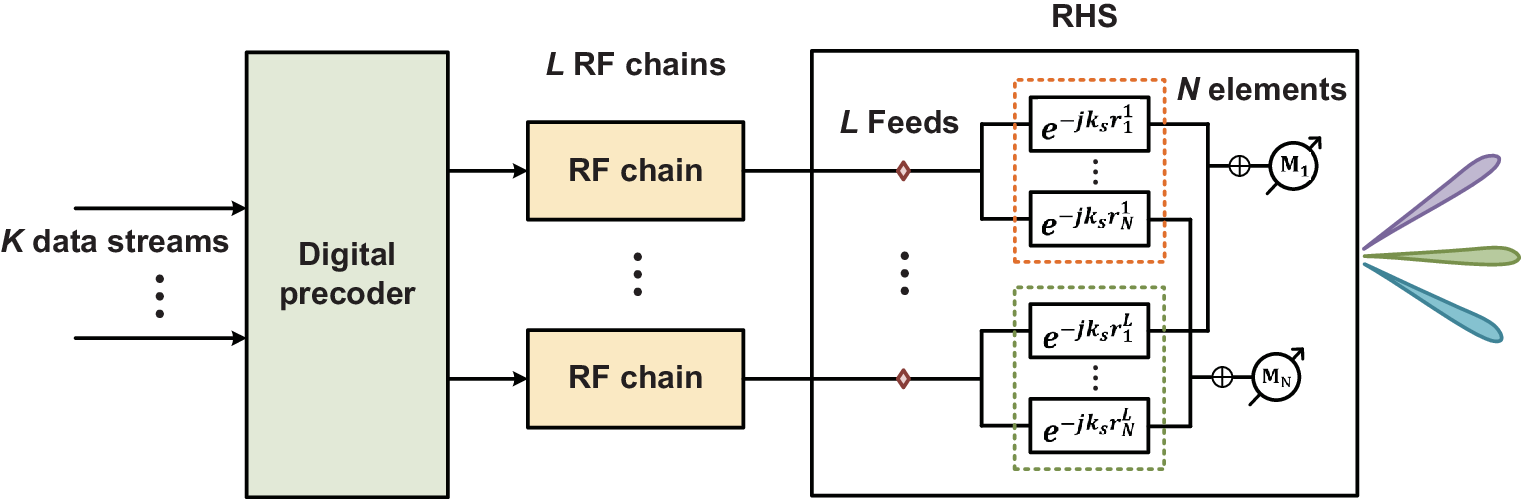}
	\caption{Block diagram of the RHS-aided system.}
	\label{Fig:block}
\vspace{0em}
\end{figure}

\section{System model}~\label{Sec:model}

In this section, a model for near-far field communication assisted by RHS is first introduced, followed by the RHS and signal models.
The sum rate maximization problem is then formulated.
\subsection{RHS-Aided Near-Far-Field Communications}

As shown in Fig.~\ref{Fig:scenario}, we consider a large-scale RHS-assisted downlink multi-user communication system, where the RHS serves as a base station~(BS) transmitter antenna array.
The BS equipped with $L$ RF chains sends $K (K\leq L)$ independent data streams to $K$ single-antenna users.
It first encodes the $K$ data streams using the digital beamformer $\mathbf{V}$, and then upconverts the signals through the RF chains. The number of RF chains equals that of RHS feeds\footnote{The feeds of RHS are integrated on the metasurface, and after signal input, EM waves propagate along the metasurface and excite RHS elements one by one.}, with each RF chain connected to one RHS feed~\cite{HDMA}. Subsequently, the feeds input the RF signals into the RHS to be radiated in free space~\cite{holoprin}. The RF chains and feeds are connected via wired connections, where the signal attenuation can be neglected~\cite{imp}.
The system diagram is shown in Fig.~\ref{Fig:block}.

Owing to the simple feeding network and low hardware cost, RHSs can be easily extended to a large-scale array.
As the scale of RHS expands, the corresponding near-field region also extends. Take the Rayleigh distance as an example, i.e., $\frac{2D^2}{\lambda}$, where $D$ represents the antenna aperture~\cite{approximate}. When the number of RHS elements reaches hundreds or thousands, its near-field region extends to tens of meters. For instance, the near-field region of a 256-element RHS at 30 GHz is 81.28 m long. In this case, users are randomly distributed both in the near field and far field of the RHS, leading to hybrid near-far-field communications.

\subsection{RHS Model}
\begin{figure}[t]
\setlength{\abovecaptionskip}{0pt}
\setlength{\belowcaptionskip}{0pt}
	\centering
    \includegraphics[width=0.43\textwidth]{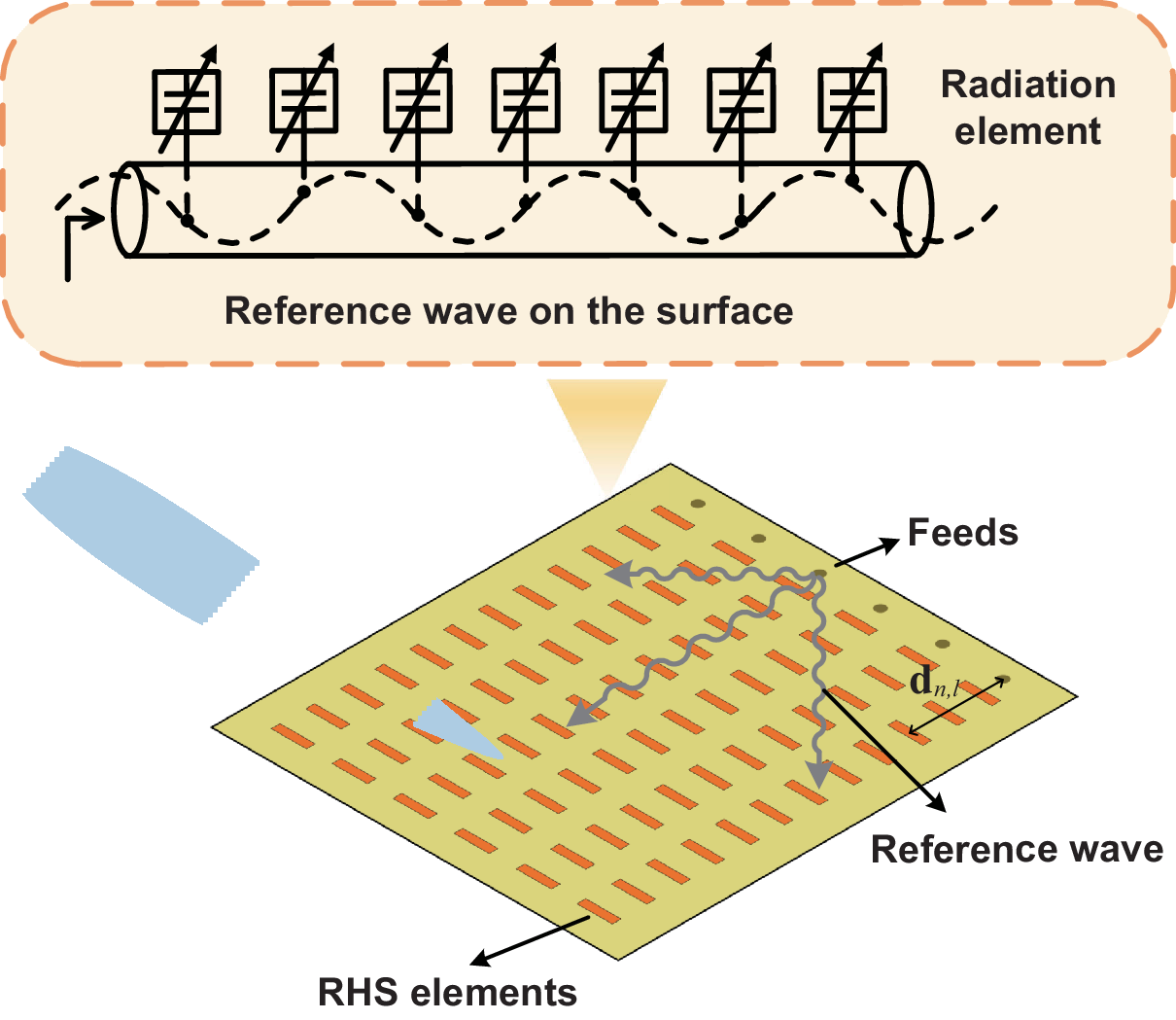}
	\caption{Illustration of the RHS.}
	\label{Fig:RHS}
\vspace{0em}
\end{figure}

As a type of leaky-wave antenna, EM waves continuously propagate along the RHS to generate directional beams as shown in Fig~\ref{Fig:RHS}. The transmit signal from the RF chain is fed into the RHS through the feeds, and the signal excites elements along the RHS sequentially~\cite{holoprin}. The response of the $l$-th feed at the $n$-th element is $e^{-j \mathbf{k}_s\cdot \mathbf{d}_{n,l}}$, where $\mathbf{d}_{n,l}$ is the distance vector between them, and $\mathbf{k}_s$ is the EM wave propagation factor of RHS. To characterize the propagating loss of the EN wave on the RHS, the path loss is represented as $e^{-\alpha |\mathbf{d}_{n,l}|}$, where $\alpha$ is the path loss factor. 
Define $\mathbf{M}\in\mathbb{C}^{N \times L}$ as the RHS holographic beamformer, where $[\mathbf{M}]_{n,l}$ is the response of the $l$-th feed at the $n$-th RHS element.
Assuming the amplitude response of the element is $m_n \in [0,1] $, the response of the $n$-th element is the sum of the EM waves from all the feeds, expressed by~\cite{imp}
\begin{equation}
M_n=\sum_{l=1}^L[\mathbf{M}]_{n,l}=\sum_{l=1}^L m_n\cdot e^{-\alpha|\mathbf{d}_{n,l}|}\cdot e^{-j\mathbf{k}_s\cdot\mathbf{d}_{n,l}}.
\end{equation}
Define the RHS amplitude vector as $\mathbf{m}=[m_1,\cdots,m_N]$. Directional beamforming can be achieved by regulating $\mathbf{m}$~\cite{holoprin}.
A configuration for $\mathbf{m}$, namely a \emph{codeword}, corresponds to a directional holographic pattern, also known as the beam pattern.

\subsection{Signal Model}
We set the center of RHS as the origin of the coordinate system, the distance between user $k$ and the origin as $r_k$, and the azimuth angle as $\theta_k$. The coordinates of the $n$-th RHS element is $(0,\delta_n d)$ where $\delta_n=\frac{2n-N-1}{2}(n=1,\cdots,N)$ and $d$ represents the antenna element spacing.
The distance from user $k$ to the $n$-th RHS element can be expressed as
\begin{align}
r_{k,n}&=\sqrt{r_k^2-2\delta_n d r_k\cos \theta_k + \delta_n^2 d^2}\\
&\stackrel{(a)}{\approx} r_k- \delta_n d \cos \theta_k +\frac{\delta_n^2 d^2(1-\cos^2\theta_k)}{2r_k}   \\
&\stackrel{(b)}{=} r_k- \delta_n d \psi_k + \frac{\delta_n^2 d^2 \mu_k}{2}, \label{psi_mu}
\end{align}
where the approximation $(a)$ is based on a Taylor expansion, and the transformation $(b)$ is the coordinate transformation via defining $\psi_k=\cos\theta_k $ and $\mu_k=\frac{1-\cos^2\theta_k}{r_k} $. 
Therefore, the channel $\mathbf{h}_k\in\mathbb{C}^{1 \times N}$ from user $k$ to the RHS can be represented as~\cite{twostage2},\cite{2D}
\begin{equation}
\mathbf{h}_k=\beta_k \sqrt{N}\mathbf{b}(\psi_k,\mu_k),
\end{equation}
where $\beta_k$ is the complex path gain. and the steering vector $\mathbf{b}(\psi_k,\mu_k)$ is expressed as
\begin{equation}
\mathbf{b}(\psi_k,\mu_k)=\frac{1}{\sqrt{N}}[e^{-j\frac{2\pi}{\lambda}r_{k,1}},\cdots,e^{-j\frac{2\pi}{\lambda}r_{k,N}}],
\end{equation}
where $r_{k,n}$ is given in~\eqref{psi_mu}, related to $\psi_k$ and $\mu_k$.

Given the channel $\mathbf{h}_k$, the digital beamformer $\mathbf{V}$, and the RHS holographic beamformer $\mathbf{M}$, the signal received by user $k$ is
\begin{equation}
y_k=\mathbf{h}_k\mathbf{M}\mathbf{v}_ks_k+\sum_{k'\neq k}\mathbf{h}_k\mathbf{M}\mathbf{v}_k's_k'+n_k,
\end{equation}
where $s_k$ is the transmitted symbol, $n\sim\mathcal{CN}\left(0,\sigma^2\right)$ is the additive white Gaussian noise, and $\mathbf{v}_k$ is the $k$-th column of the digital beamformer $\mathbf{V}$.
The data rate of each user is determined by the received signal power, inter-user interference, and noise, which is given by~\cite{sumrate}
\begin{equation}
R_k=\log_2\left(1+\frac{|\mathbf{h}_k\mathbf{M}\mathbf{v}_k|^2}{\sigma^2+\sum_{k'\neq k}|\mathbf{h}_k\mathbf{M}\mathbf{v}_k'|^2}\right).
\end{equation}

\subsection{Problem Formulation}
In the RHS-aided multi-user system,  we aim to maximize the sum data rate of users through the design of the digital beamformer $\mathbf{V}$ and the holographic beamformer $\mathbf{M}$.
This problem can be formulated as
\begin{subequations}\label{problem}
\begin{align}
\max_{\mathbf{V},\mathbf{m}}\quad \sum_{k=1}^K& \log_2\left(1+\frac{|\mathbf{h}_k\mathbf{M}\mathbf{v}_k|^2}{\sigma^2+\sum_{k'\neq k}|\mathbf{h}_k\mathbf{M}\mathbf{v}_k'|^2}\right) \\
\text{s.t.}\text{ } &Tr(\mathbf{Vs} \mathbf{s}^H\mathbf{V}^H)= P,\label{con1}\\
     &0\leq\mathrm{m}_{n}\leq1,\quad\forall n,\label{con2}\\
      &\eta Tr(\mathbf{MVs} \mathbf{s}^H\mathbf{V}^H\mathbf{M}^H) \leq P,\label{con3}
\end{align}
\end{subequations}
where $\eta$ is the energy efficiency of the RHS element, defined as the ratio of the energy radiated by each element to the total input energy.
Constraint~\eqref{con1} is the power constraint for the BS, $P$ is the power budget available at the BS, and constraint~\eqref{con2} is the available amplitude range.
\eqref{con3} is the leaky power constraint, which means that the energy input from the feeds of the RHS gradually decreases after passing through each element.

Due to the coupling between holographic beamformer $\mathbf{M}$ and digital beamformer $\mathbf{V}$, obtaining the optimal solution to Problem~\eqref{problem} is extremely challenging~\cite{multiuser1}. To address this, we consider a low-complexity approach to design the hybrid beamformer, which involves first configuring $\mathbf{M}$ to direct beams towards $K$ users, and then designing $\mathbf{V}$ to eliminate inter-user interference.
For the holographic beamformer $\mathbf{M}$, we consider a practical codebook-based approach where the optimal holographic pattern is selected from a pre-designed codebook $\mathcal{M}$ to configure the RHS beamformer $\mathbf{M}$ via beam training~\cite{multiuser2}. After that, the design of $\mathbf{V}$ is accomplished using general methods like the zero-forcing~(ZF) algorithm. 

We mainly focus on optimizing the holographic beamformer through codebook design and beam training. Specifically, we configure the RHS using the codeword $\mathbf{m}$ from a codebook $\mathcal{M}$ to maximize the beamforming gain for $K$ users, expressed as~\cite{far_multi-user}
\begin{subequations}\label{problem_bt}
\begin{align}
\max_{\mathbf{m}}&\quad \sum_{k=1}^K |\mathbf{b}(\psi_k,\mu_k)\mathbf{M}\mathbf{v}_k|^2 \\
\text{s.t.}\text{ } &Tr(\mathbf{V}^H\mathbf{V})= P, \\
&\mathbf{m}\in \mathcal{M},
\end{align}
\end{subequations}
where the user channel vector $\mathbf{b}$ is unknown and $\mathbf{v}_k$ is initialized as a constant vector in the beam training.
This is a typical design problem for a codebook and beam training scheme. Subject to the constraints in \eqref{con2} and \eqref{con3}, the initial step involves designing the codebook structure, the number of codewords, and the beam patterns of these codewords. Subsequently, an optimal codeword is selected by devising a beam training scheme, which entails a process for searching through the codewords.

We aim to search the optimal holographic pattern $\mathbf{m}$ through the codebook in a low-cost manner, the multi-lobe beam range of which can simultaneously cover all $K$ users.
Due to the hybrid near-far field feature and the unique holographic pattern design of RHSs, problem~\eqref{problem_bt} is non-trivial to solve.
\begin{itemize}
\item First, since the user channel $\mathbf{b}(\psi_k,\mu_k)$ is related to both distance and angle, beam training is required to be performed in the distance-angle binary domain~\cite{exh}. Traditional beam training approaches such as exhaustive search and two-phase beam search face significant overhead. For instance, in a system with 512 angle samplings and 10 distance samplings, the overhead of these two schemes are 5120 and 522, respectively. It is crucial to design low-overhead beam training schemes to enhance effective communication time and system throughput.
\item Second, most traditional codebooks and beam training schemes are based on parallel-fed phased arrays or RISs that utilize phase modulation devices.
In contrast, RHSs as a type of series-fed antenna array, exhibit the benefit of holographic pattern additivity subject to the leaky power constraint~\cite{HDMA}.
That is, the superposition of pattern $\mathbf{m}_1$ and pattern $\mathbf{m}_2$ can directly generate a new pattern.
It necessitates the design of a tailored holographic codebook-based beam training scheme.
\item Third, users are distributed both in the near and far fields in large-scale array systems in practical, i.e., $\psi_k$ and $\mu_k$ all across the near-far field~\cite{Holo2}.
Traditional pure near- or far-field beam training schemes are typically based on spherical or planar wave channels respectively, making it difficult to be universally applicable for both near- and far-field users. Therefore, it is crucial to design beam training schemes that are applicable to both near-field and far-field users.

\end{itemize}

\section{Framework of RHS Multi-user Codebook and Beam Training}\label{Sec:framework}

In this section, the design objectives for the multi-user codebook and beam training schemes are first presented. Subsequently, a framework for codebook and beam training is proposed.

\subsection{Design Objectives of RHS Codebook-based Beam Training}
When users are distributed in both the near and far fields, we aim to find the optimal codewords (defined by \eqref{problem_bt}) for all users concurrently with low overhead, through the design of the codebook $\mathcal{M}$ and the beam training scheme.
The RHS-assisted multi-user codebook-based beam training scheme is expected to exhibit the following characteristics.
\begin{itemize}
\item \textbf{Low training overhead}: In contrast to traditional near-field beam training schemes where the training overhead scales linearly with the number of antennas, the proposed scheme requires low time overhead even in large-scale systems.
\item \textbf{One-shot multi-user training}: Unlike traditional beam training methods that serve users individually, the designed beam training scheme utilizes RHS codebooks to enable one-shot training of all users in the system, further reducing the total beam training overhead.
\item \textbf{Near-far-field adaptability}: Unlike traditional codebook-based beam training schemes that are limited to either near-field or far-field users, the designed scheme is applicable to both near-field and far-field users.
\item \textbf{Holographic-pattern superposition}: In contrast to traditional codebooks and beam training methods based on phased arrays or phase-shifting metasurfaces, RHS codebooks and beam training schemes utilize the superposition of holographic patterns to generate real-time adaptive codewords givin the leaky power constraint.
\end{itemize}

\subsection{Structure of RHS Multi-user Beam Training and Codebooks}

\begin{figure}[t]
\setlength{\abovecaptionskip}{0pt}
\setlength{\belowcaptionskip}{0pt}
	\centering
    \includegraphics[width=0.45\textwidth]{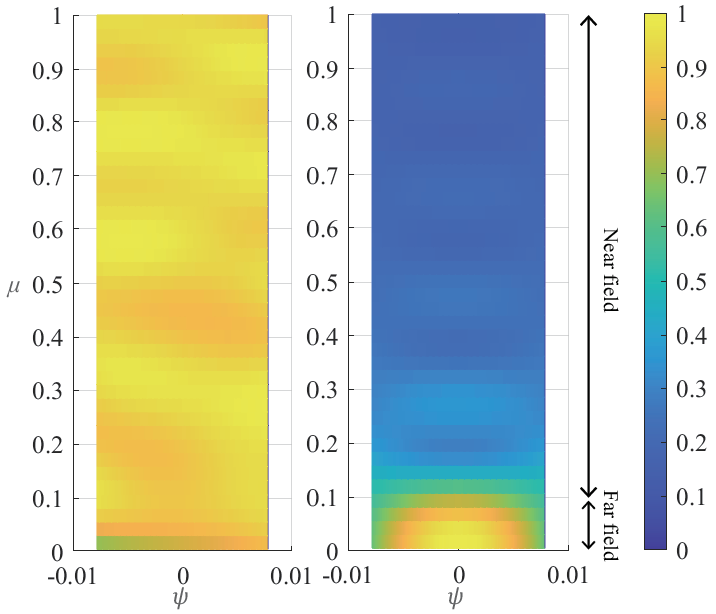}
	\caption{Near-far-field angular codeword vs. traditional far-field codeword.}
	\label{Fig:near_far_codeword}
\vspace{0em}
\end{figure}

We design the beam training process in two phases, each utilizing a customized codebook. In Section~\ref{Sec:model}, we represent the channels in the $\psi-\mu$ domain as shown in~\eqref{psi_mu}, which is equivalent to the $r-\theta$ domain\footnote{In the $\psi-\mu$ domain, given the angle $\psi = \cos\theta$, the parameter $\mu$ decreases as the distance $r$ increases. This implies that $\mu$ in the far-field region is smaller than that in the near-field region, as illustrated in Fig.~\ref{Fig:near_far_codeword}.}. Below, we introduce the beam training process in the $\psi-\mu$ domain.

\subsubsection{\textbf{Simultaneous multi-user angle search}}
\begin{figure}[t]
\setlength{\abovecaptionskip}{0pt}
\setlength{\belowcaptionskip}{0pt}
	\centering
    \includegraphics[width=0.45\textwidth]{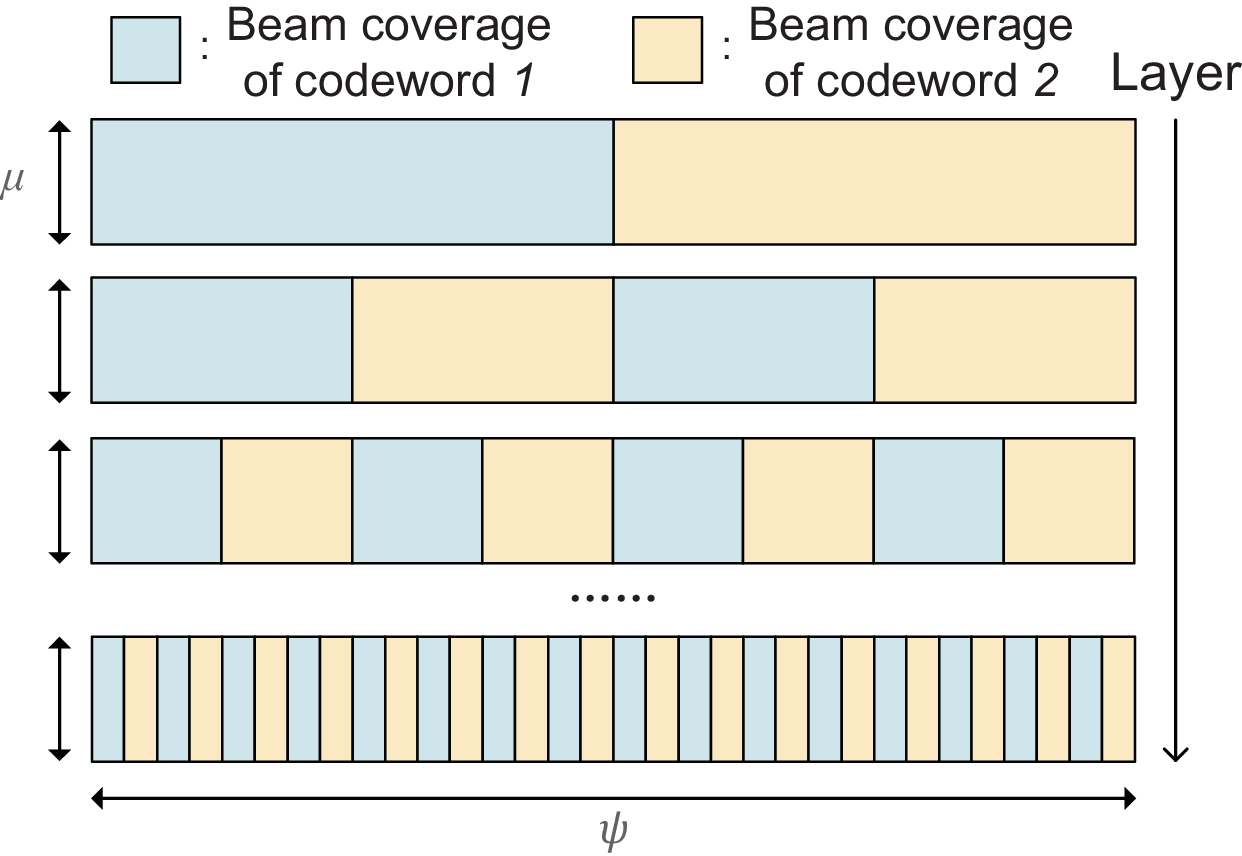}
	\caption{Structure of RHS multi-user angular codebook.}
	\label{Fig:angularbook}
\vspace{0em}
\end{figure}

To achieve a fast angle search for all users in one shot, we propose a hierarchical codebook as shown in Fig.~\ref{Fig:angularbook}.
Each layer consists of two codewords, where each codeword represents a multi-lobe beam, covering specific discontinuous angles.
As shown in Fig.~\ref{Fig:near_far_codeword}, a codeword corresponding to a specific angle exhibits high beamforming gain no matter what the distance is. This is different from the traditional far-field angular codewords that only achieve a high beamforming gain in the far field.
During beam training process, two codewords from each layer are successively used to transmit signals and obtain feedback from all users. Based on the coverage relationship of codewords in adjacent layers, all the user angles can be calculated utilizing the feedback from each layer.
The detailed design of the multi-user angular codebook is discussed in Section~\ref{Sec:angular_book}.
\subsubsection{\textbf{Simultaneous multi-user distance search}}
\begin{figure}[t]
\setlength{\abovecaptionskip}{0pt}
\setlength{\belowcaptionskip}{0pt}
	\centering
    \includegraphics[width=0.45\textwidth]{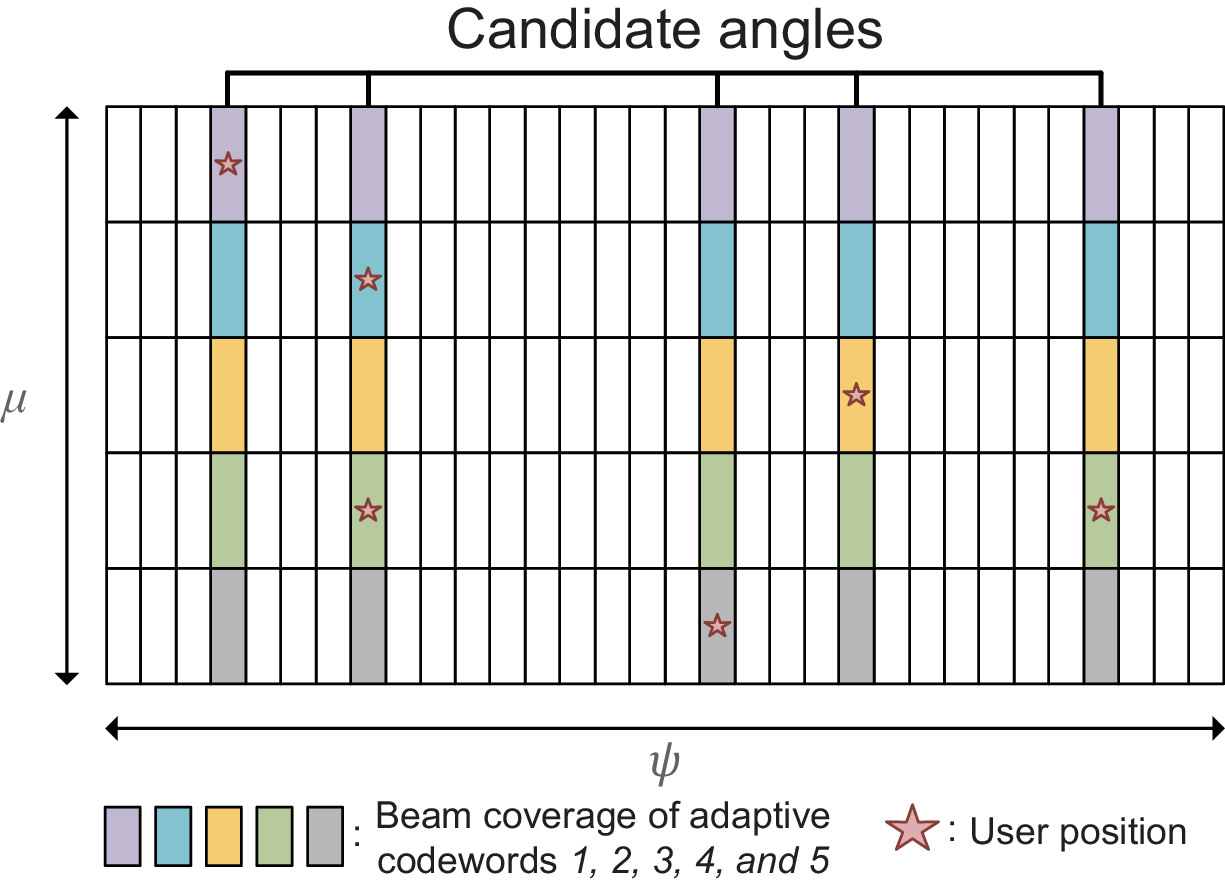}
	\caption{Structure of RHS distance-adaptive codebook.}
	\label{Fig:rangebook}
\vspace{0em}
\end{figure}

After the simultaneous angle search, the BS identifies the potential angles for all users, and then performs a distance search for all users simultaneously. As shown in Fig.~\ref{Fig:rangebook}, assuming users are located at the stars, after the angle search, the candidate angles for all users are determined. We aim to design distance-adaptive codewords, where each multi-lobe codeword covers a specific $\mu$-distance for all candidate angles. By using adaptive codewords for beam training one by one, users feed back the codeword indices that maximize the received signal power, thereby the distance search is completed simultaneously. Details on real-time and fast generation of distance-adaptive codewords based on holographic principles are shown in Section~\ref{Sec:distance_book}.

\section{RHS Multi-user Angular Codebook Design}~\label{Sec:angular_book}
In this section, the design techniques for multi-beam codewords in the angular domain and the multi-user angular codebook as shown in Fig.~\ref{Fig:angularbook} are introduced.

\subsection{RHS Multi-user Angular Codeword Design}~\label{subec:codeword}
Suppose that the $p$-th codeword in the $s$-th layer is $\mathbf{m}_{s,p}$ and the desired coverage of the codebook is $\psi \in [\psi_{min},\psi_{max}]$ and $\mu \in [\mu_{min},\mu_{max}]$.
To design codewords that exhibit high beamforming gain in multiple areas, we conduct $I$ and $J$ equally spaced samplings in the $\psi$ and $\mu$ domains, respectively. The set of sample points is $\mathcal{S}=[(\psi_1, \mu_1),\cdots,(\psi_i, \mu_j),\cdots,(\psi_I, \mu_J))]$. $(\psi_i, \mu_j)$ is represented as $\psi_i=\psi_{min}+\frac{\psi_{max}-\psi_{min}}{2I}\cdot(2i-1)$ and $\mu_j=\mu_{min}+\frac{\mu_{max}-\mu_{min}}{2J}\cdot(2j-1)$ where $i\in \{1,2,\cdots,I\}$ and $j\in \{1,2,\cdots,J\}$.
Define the beam coverage of codeword $\mathbf{m}_{s,p}$ as $\mathcal{C}(\mathbf{m}_{s,p})$,
the point $(\psi_i, \mu_j)$ in $\mathcal{C}(\mathbf{m}_{s,p})$ satisfy
\begin{equation}\label{coverage}
\lceil \frac{i\cdot2^s}{I} \rceil \bmod 2 = p.
\end{equation}

To achieve high beamforming gain for codeword $\mathbf{m}_{s,p}$ in its coverage $\mathcal{C}(\mathbf{m}_{s,p})$, and almost zero beamforming gain in other areas, the codeword design problem of $\mathbf{m}_{s,p}$ can be expressed as
\begin{equation}\label{P_code}
\begin{aligned}
&\min_{\mathbf{m}}\quad \sum_{i=1}^I \sum_{j=1}^J (|\mathbf{b}(\psi_i,\mu_j)\mathbf{M}\mathbf{v}|- G_{i,j})^2 \\
&\begin{array}{r@{\quad}r@{}l@{\quad}l}
\mathbf{s.t.}
&G_{i,j}=
&\begin{cases}
 D  ,\text{ if } (\psi_i, \mu_j) \in \mathcal{C}(\mathbf{m}_{s,p}), \\
 0,  \text{ else}.
\end{cases}\\
\\
&  \quad 0\leq\mathrm{m}&_{n}\leq1,\quad\forall n,\\
\end{array}
\end{aligned}
\end{equation}
where the beamforming gain $D$ is a constant, and the digital beamformer is initialized as $\mathbf{v}= \sqrt{\frac{P}{LK}}[1,\cdots,1]^H$~\cite{digital}, since there is no need to eliminate inner-user interference during simultaneous multi-user beam training.
The optimization term of the codeword at sample point $(\psi_i, \mu_j)$ is
\begin{equation}
(|\mathbf{b}(\psi_i,\mu_j)\widetilde{\mathbf{m}}|- G_{i,j})^2
\end{equation}
where $\widetilde{\mathbf{m}}=\mathbf{M}\mathbf{v}$ and $[\widetilde{\mathbf{m}}]_{n}=\sum_{l=1}^L\sqrt{\frac{\eta P}{LK}}\cdot m_n\cdot e^{-\alpha|\mathbf{d}_{n,l}|}\cdot e^{-j\mathbf{k}_s\cdot\mathbf{d}_{n,l}}$.
Let $\gamma_n=\sum_{l=1}^L\sqrt{\frac{\eta P}{LK}}\cdot e^{-\alpha|\mathbf{d}_{n,l}|}\cdot e^{-j\mathbf{k}_s\cdot\mathbf{d}_{n,l}}$, the $\widetilde{\mathbf{m}}$ can be expressed by $[\gamma_1 m_1,\cdots,\gamma_N m_N]$.
The following two cases are discussed: $G_{i,j}=0$ or $G_{i,j}=D$.

\begin{figure}[t]
\setlength{\abovecaptionskip}{0pt}
\setlength{\belowcaptionskip}{0pt}
	\centering
    \includegraphics[width=0.45\textwidth]{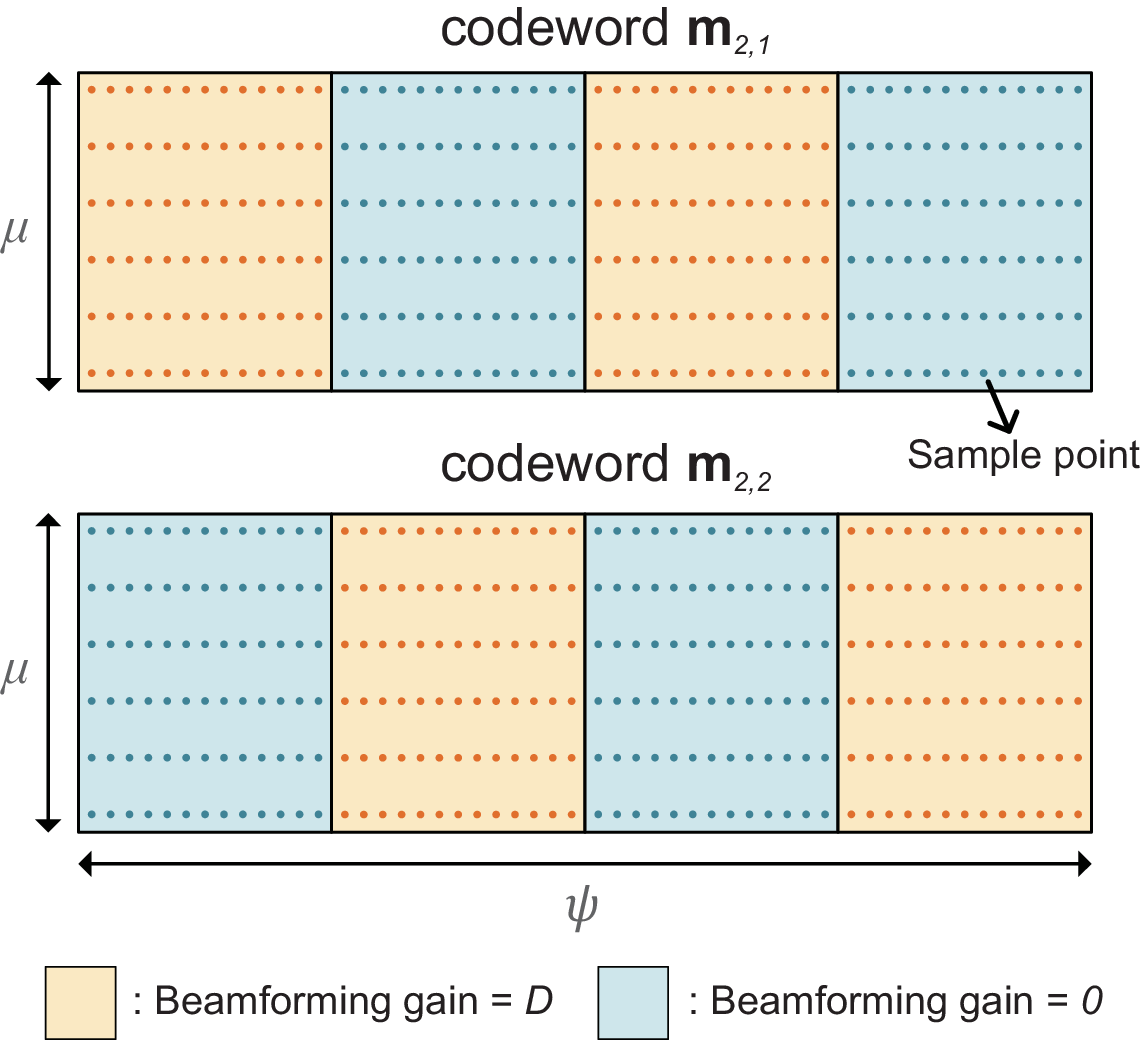}
	\caption{Desired beamforming gain of $\mathbf{m}_{2,1}$ and $\mathbf{m}_{2,2}$.}
	\label{Fig:codeword}
\vspace{0em}
\end{figure}

\textbf{Case 1:}  When $(\psi_i,\mu_j) \notin \mathcal{C}(\mathbf{m}_{s,p})$, that is, located in the blue area in Fig.~\ref{Fig:codeword} and $G_{i,j}=0$, the optimization term of the codeword at sample point $(\psi_i, \mu_j)$ can be expressed as
\begin{equation}
\begin{aligned}
&|\mathbf{b}(\psi_i,\mu_j)\widetilde{\mathbf{m}}|^2 \\
=& \widetilde{\mathbf{m}}^H \mathbf{b}(\psi_i,\mu_j)^H\mathbf{b}(\psi_i,\mu_j)\widetilde{\mathbf{m}} \\
=& \sum_{n_1=1}^{N}\sum_{n_2=1}^{N} \gamma_{n_1}^H\gamma_{n_2}m_{n_1}m_{n_2}[\mathbf{B}]_{n_1,n_2}
\end{aligned}
\end{equation}
where $\mathbf{B}=\mathbf{b}(\psi_i,\mu_j)^H\mathbf{b}(\psi_i,\mu_j)$.
This is a quadratic function with respect to $m_n$ when fixing the other $m_n'(n'\neq n)$, which can be expressed as
\begin{equation}\label{eq15}
\begin{aligned}
&|\mathbf{b}(\psi_i,\mu_j)\widetilde{\mathbf{m}}|^2 \\
=& a_{i,j,n}m_n^2 + b_{i,j,n}m_n+ c_{i,j,n},
\end{aligned}
\end{equation}
where the coefficients $a_{i,j,n}$, $b_{i,j,n}$, and $c_{i,j,n}$ of this quadratic function are represented as 
\begin{equation}\label{case1}
\begin{aligned}
&a_{i,j,n}=|\gamma_{n}|^2[\mathbf{B}]_{n,n} \\
&b_{i,j,n}=2\text{Re}\{{\sum_{n'\neq n}^{N}\gamma_n^H\gamma_{n'}[\mathbf{B}]_{n,n'} m_{n'}}\} \\
&c_{i,j,n}=2\text{Re}\{ \sum_{n_1\neq n}^{N}\sum_{n_2=1}^{n1} \gamma_{n_1}^H\gamma_{n_2}m_{n_1}m_{n_2}[\mathbf{B}]_{n_1,n_2} \}.
\end{aligned}
\end{equation}

\textbf{Case 2:}  When $(\psi_i,\mu_j) \in \mathcal{C}(\mathbf{m}_{s,p})$, that is, located in the yellow area in Fig.~\ref{Fig:codeword} and $G_{i,j}=D$, the optimization term of the codeword at sample point $(\psi_i, \mu_j)$ can be expressed as
\begin{equation}\label{case2}
\begin{aligned}
&(|\mathbf{b}(\psi_i,\mu_j)\widetilde{\mathbf{m}}|-D)^2 \\
&= |\mathbf{b}(\psi_i,\mu_j)\widetilde{\mathbf{m}}|^2-2D|\mathbf{b}(\psi_i,\mu_j)\widetilde{\mathbf{m}}|+ D^2 \\
&= a_{i,j,n}m_n^2 + b_{i,j,n}m_n+ c_{i,j,n}.
\end{aligned}
\end{equation}
Through the third-order Taylor expansion of \eqref{eq15} at $x=0$, the coefficients $a_{i,j,n}$, $b_{i,j,n}$, and $c_{i,j,n}$ of \eqref{case2} are represented as 
\begin{equation}\label{case2_}
\begin{aligned}
&a_{i,j,n}=a_0-\frac{(4a_0c_0-b_0^2)D}{4c^{\frac{3}{2}}},\\
&b_{i,j,n}=b_0-\frac{b_0D}{c_0^{\frac{1}{2}}},\\
&c_{i,j,n}=c_0-2Dc_0^\frac{1}{2}+D^2,
\end{aligned}
\end{equation}
where the values of $a_0$, $b_0$, and $c_0$ are $a_{i,j,n}$, $b_{i,j,n}$, and $c_{i,j,n}$ in \eqref{case1}, respectively.

To solve the minimization problem in \eqref{P_code}, we iteratively optimize $\mathbf{m}$. For each $m_n$, when fixing other $m_n'(n'\neq n)$, the term corresponding to all the sample points can be expanded into a quadratic function about $m_n$ as
\begin{equation}\label{quar_fun}
\mathbf{f}(m_n)=a_{n}m_n^2 + b_{n}m_n+ c_{n},
\end{equation}
where $a_{n}=\sum_{i=1}^{I}\sum_{j=1}^{J}a_{i,j,n}, b_{n}=\sum_{i=1}^{I}\sum_{j=1}^{J}b_{i,j,n}, c_{n}=\sum_{i=1}^{I}\sum_{j=1}^{J}c_{i,j,n}.$
Therefore, the minimum of this function on $[0,1]$ is found by the properties of the quadratic function as below.
\begin{itemize}
\item \textbf{$a_{n} \leq 0$}: If $-\frac{b_n}{2a_n} \leq \frac{1}{2}$, $m_n=1$.
If $-\frac{b_n}{2a_n} \textgreater \frac{1}{2}$, $m_n=0$.
\item \textbf{$a_{n} \textgreater 0$}: If $-\frac{b_n}{2a_n} \leq \frac{1}{2}$, $m_n=\text{max}(-\frac{b_n}{2a_n},0)$.
If $-\frac{b_n}{2a_n} \textgreater \frac{1}{2}$, $m_n=\text{min}(-\frac{b_n}{2a_n},1)$.
\end{itemize}

When optimizing $m_n$ iteratively, since the value in \eqref{P_code} is reduced during each iteration, convergence is achieved.
The design method of RHS multi-user angular codewords is summarized in Algorithm~\ref{alg:codeword}.

\begin{algorithm}[t]
\label{alg:codeword}
\caption{RHS Multi-user Angular Codeword Design}
\LinesNumbered
\KwIn{The set of sample points $\mathcal{S}$, codeword coverage $\mathcal{C}(\mathbf{m}_{s,p})$, desired beamforming gain $D$.}

\Repeat{convergence}
{
      \For {$n=1:N$}
{
Calculate the coefficients $a_{n}$, $b_{n}$, and $c_{n}$ in~\eqref{quar_fun};\\
\eIf{$a_{n} \leq 0$}{$m_n \leftarrow$ the value farthest from $-\frac{b_n}{2a_n}$ in $[0,1]$;}{$m_n \leftarrow$ the value closest to $-\frac{b_n}{2a_n}$ in $[0,1]$;}
}
}
\KwOut{RHS multi-user angular codeword $\mathbf{m}_{s,p}$.}
\end{algorithm}

\subsection{RHS Multi-User Angular Codebook Design}

After providing the design scheme of multi-user angular codewords, we now present the generation method of the RHS multi-user angle codebook. In the hierarchical codebook, the coverage of the codewords remains constant, but the number of beams within the codewords gradually increases with each layer, and the beam width gradually narrows. Given the RHS array, the width of the beams generated by the system can not be narrowed continuously, that is, the number of layers of the hierarchical codebook has an upper limit. The number of codebook layers should ensure that the theoretical beam width of the bottom-layer codeword is not less than the narrowest beam the system can produce, represented as~\cite{beamwidth}
\begin{equation}\label{bottom}
\frac{\psi_{max}-\psi_{min}}{2^S} \leq \frac{\lambda}{Nd},
\end{equation}
where $S$ represents the layer of the codebook, and $Nd$ is the antenna aperture.
Given the codebook coverage, the number of codebook layers, and the set of sampling points, the RHS multi-user angular codebook can be designed layer by layer through the codeword generation method in subsection~\ref{subec:codeword}, which is summarized in Algorithm 2.

\section{RHS Multi-user Distance-Adaptive Codebook Design}\label{Sec:distance_book}

Now we introduce the design method of RHS multi-user distance-adaptive codebook shown in Fig.~\ref{Fig:rangebook} to support the second phase of multi-user beam training, which is based on the superposition of holographic patterns. After the first phase of beam training in the RHS angular codebook, the BS obtains the angle information of each user. The design goal of the distance-adaptive codebook is to dynamically generate a distance-domain codebook that only covers the user angle with low overhead. 

\subsection{Limitations of PA/RIS in Distance-Adaptive Codeword Design}
The superposition of single-beam codewords is expected to achieve the design of multi-beam codewords, which can quickly generate distance-adaptive codewords during beam training.
For traditional phase-controlled arrays and metasurfaces, it is difficult to achieve this scheme, as explained in the following two cases.
\begin{itemize}
\item \textbf{Direct superposition of single-beam codewords:}
Assume that there are codewords $[\mathbf{b}_1,\cdots,\mathbf{b}_N]$ aligned with coordinates $[(\psi_1,\mu_1),\cdots,(\psi_N,\mu_N)]$ respectively. Through the direct superposition of codewords, that is, $\frac{1}{N}\sum_{n=1}^N \mathbf{b}_n$, it is unachievable to generate a multi-beam codeword covering $[(\psi_1,\mu_1),\cdots,(\psi_N,\mu_N)]$ at the same time.
Due to the use of analog beamformers in multi-user beam training, for phased arrays and metasurfaces like RISs, the response of the elements is constrained by constant modulus and cannot achieve direct superposition of codewords~\cite{modulus}.

\item \textbf{Optimized multi-beam codeword:}
By optimizing the weight $[e^{j\theta_1},\cdots,e^{j\theta_N}]$ of the codeword, a multi-beam codeword covering $[(\psi_1,\mu_1),\cdots,(\psi_N,\mu_N)]$ simultaneously can be generated, which satisfies the constant modulus constraint of the elements. However, this optimization scheme requires a large amount of time overhead~\cite{modulus}, making it difficult to dynamically generate multi-beam codewords in real-time during beam training.
\end{itemize}

\begin{algorithm}[t]
\label{alg:codebook}
\caption{RHS Multi-user Angular Codebook Design}
\LinesNumbered
\KwIn{The set of sample points $\mathcal{S}$, codebook coverage $[\psi_{min},\psi_{max}]$ and $[\mu_{min},\mu_{max}]$ , codebook layer $S$.}

      \For {$s=1:S$}
{
\For {$p=1:2$}
{
Calculate the codeword coverage $\mathcal{C}(\mathbf{m}_{s,p})$ according to \eqref{coverage};\\
Generate the codeword $\mathbf{m}_{s,p}$ based on Algorithm~\ref{alg:codeword};\\
Add $\mathbf{m}_{s,p}$ into $\mathcal{M}_{angle}$;
}
}
\KwOut{RHS multi-user angular codebook $\mathcal{M}_{angle}$.}
\end{algorithm}

\subsection{RHS Multi-user Distance-Adaptive Codeword Design}
For RHS, the superposition of single-beam codewords, i.e. holographic patterns, can be achieved through holographic principles to generate multi-beam codewords.
During the superposition process of the RHS single-beam pattern, the amplitude of each element is superposed while the phase information remains unchanged, thereby avoiding the aliasing phenomenon in the beam pattern.

To generate multi-beam codewords that avoid holographic pattern aliasing after superposition, we first introduce the design methodology of RHS single-beam pattern that is aligned at $(\psi_i,\mu_i)$. The design objective of this single-beam pattern is to make the energy converge on $(\psi_i,\mu_i)$ as much as possible, which we formulate as
\begin{subequations}
\begin{align}
\max_{\mathbf{m}}&\quad |\mathbf{b}(\psi_i,\mu_j)\mathbf{M}\mathbf{v}|\\
\text{s.t.}\text{ } &\quad 0\leq\mathrm{m}_{n}\leq1,\quad\forall n,\\
&\eta Tr(\mathbf{MVs} \mathbf{s}^H\mathbf{V}^H\mathbf{M}^H) \leq P,
\end{align}
\end{subequations}
where the digital beamformer is initialized as $\mathbf{v}= \sqrt{\frac{P}{LK}}[1,\cdots,1]^H$~\cite{digital}.
This is a typical RHS holographic beamforming problem, which is solved in~\cite{imp} and summarized as Algorithm 3.

\begin{algorithm}[t]
\label{alg:single_codeword}
\caption{RHS Single-Beam Pattern Design}
\LinesNumbered
\KwIn{Target coordinate $(\psi_i, \mu_j)$, RHS element number $N$, initialized digital beamformer $\mathbf{v}$.}
\Repeat{convergence}
{
      \For {$n=1:N$}
{Set $m_n=0$, compute $S_0=|\mathbf{b}(\psi_i,\mu_j)\mathbf{M}\mathbf{v}|$; \\
Set $m_n=1$, compute $S_1=|\mathbf{b}(\psi_i,\mu_j)\mathbf{M}\mathbf{v}|$;  \\
\eIf{$S_0 \textgreater S_1$}{$m_n \leftarrow 0$;}{$m_n \leftarrow 1$;}
}
}
\KwOut{RHS single-beam pattern $\mathbf{m}$.}

\end{algorithm}

Assuming that single-beam patterns $[\mathbf{m}_1,\cdots,\mathbf{m}_K]$ aligned with coordinates $[(\psi_1,\mu_1),\cdots,(\psi_K,\mu_K)]$ respectively are generated using Algorithm~\ref{alg:single_codeword}, by leveraging the amplitude-control characteristics of the RHS, a multi-beam codeword $\mathbf{m}$ that simultaneously covers $[\mathbf{m}_1,\cdots,\mathbf{m}_K]$ is constructed through the superposition of these single-beam patterns, expressed as
\begin{equation}
\mathbf{m}=\frac{1}{K} \sum_{k=1}^K \mathbf{m}_K,
\end{equation}
where each element of $\mathbf{m}$ remains in $[0,1]$. The beam pattern of the multi-beam codeword is shown in Section~\ref{Sec:simulation}.

\subsection{RHS Multi-user Distance-Adaptive Codebook Design}
Since the generation method of multi-beam codewords is the direct superposition of single-beam patterns, we first design a single-beam codebook $\mathcal{M}_{singlebeam}$ in advance, which comprises all possible single-beam patterns that could be used.
Given the sample points $(\psi_i, \mu_j)$ where $i\in \{1,2,\cdots,I\}$ and $j\in \{1,2,\cdots,J\}$,
there are $I\cdot J$ singlebeam patterns.
$\mathcal{M}_{singlebeam}=[\mathbf{m}(\psi_1, \mu_1),\cdots,\mathbf{m}(\psi_I, \mu_J)]$ where the holographic pattern $\mathbf{m}(\psi_i, \mu_j)$ is generated as Algorithm~\ref{alg:single_codeword} to align with $(\psi_i,\mu_j)$.

As shown in Fig.~\ref{Fig:rangebook}, to accomplish the second phase of multi-user beam training, it is necessary to design a distance-adaptive codebook $\mathcal{M}_{distance}$ consisting of multiple multi-beam codewords, each covering a specific distance of all user angles.
Assuming that after the first phase of multi-user beam training, the angle set where the users are located is $\mathbf{\Psi}=[\psi_1, \cdots, \psi_K]$, hence the $j$-th multi-beam codeword can be designed as
\begin{equation}\label{multi-distance}
\mathbf{m}_j=\frac{1}{K} \sum_{k=1}^K \mathbf{m}(\psi_K, \mu_j).
\end{equation}
Therefore the distance-adaptive codebook $\mathcal{M}_{distance}=[\mathbf{m}_1,\cdots,\mathbf{m}_J]$. The distance-adaptive codebook design is summarized as Algorithm~\ref{alg:distance_codebook}.

\begin{algorithm}[t]
\label{alg:distance_codebook}
\caption{RHS Distance-Adaptive Codebook Design}
\LinesNumbered
\KwIn{The set of user angle $\mathbf{\Psi}$, the set of sample points $\mathcal{S}$.}
\For {$i=1:I$}
{
\For {$j=1:J$}
{
Generate single-beam codeword $\mathbf{m}(\psi_i,\mu_j)$ using Algorithm~\ref{alg:single_codeword};\\

Add $\mathbf{m}(\psi_i,\mu_j)$ into $\mathcal{M}_{singlebeam}$;
}

}

\For {$j=1:J$}
{
Generate multi-user distance-adaptive codeword $\mathbf{m}_j$ according to \eqref{multi-distance};\\
Add $\mathbf{m}_j$ into $\mathcal{M}_{distance}$;
}

\KwOut{RHS multi-user distance-adaptive codebook $\mathcal{M}_{distance}$.}

\end{algorithm}

\section{RHS-Aided Simultaneous Multi-user Beam Training and Beamformer Design}\label{Sec:beamtraining}

In this section, we first introduce a two-phase simultaneous multi-user beam training scheme, and then provide an illustrative example.
Finally, the design method of beamformers is presented.

\subsection{Two-Phase Simultaneous Beam Training}
After designing the multi-user angular codebook $\mathcal{M}_{angle}$ and the multi-user distance-adaptive codebook $\mathcal{M}_{distance}$, to solve the beam training problem in ~\eqref{problem_bt}, a two-phase simultaneous multi-user beam training utilizing the designed codebooks is proposed.

\textbf{Simultaneous angle search phase:} This stage performs beam search of the user angle layer by layer within the codebook $\mathcal{M}_{angle}$.
In the $s$-th layer, the BS configures the amplitude of the RHS element using codewords $\mathbf{m}_{s,1}$ and $\mathbf{m}_{s,2}$ respectively, while the digital beamformer remains constant as $\mathbf{v}_k= \sqrt{\frac{P}{LK}}[1,\cdots,1]^H$~\cite{digital}. User $k$ receives the signals and records the index $\tau_{k,s}$ of the codeword that results in a higher received power. Upon completing the angular search across all $S$ layers, leveraging the relationship between the codeword coverage of adjacent layers and the feedback $[\tau_{k,1},…,\tau_{k,S}]$ of user $k$  from each layer, the corresponding beam index $\nu_{k,s}$ for user $k$ in the $s$-th layer is calculated as
\begin{equation}\label{index}
\nu_{k,s}=2(\nu_{k,s-1}-1)+\tau_{k,s},
\end{equation}
where $\nu_{k,0}=1$.
The beam index $\nu_{k,s}$ of each user is fed back to BS.
Hence, the angle of user $k$ is obtained by
\begin{equation}\label{psi}
\hat{\psi_k}= \psi_{min}+\frac{\psi_{max}-\psi_{min}}{2^{S+1}}\cdot(2\nu_{k,S}-1).
\end{equation}
During the angle search process, all users take the beam training simultaneously.
The BS only needs to traverse the codewords in the codebook $\mathcal{M}_{angle}$ once to obtain the angle set $\mathbf{\Psi}=[\hat{\psi_1},\cdots,\hat{\psi_k}]$ for all users.

\textbf{Simultaneous distance search phase:} After the angle search stage, the BS acquires the set of angles $\mathbf{\Psi}$ where the users are located. To facilitate simultaneous search in the distance domain for multiple users, Algorithm~\ref{alg:distance_codebook} is first utilized to generate an RHS distance-adaptive codebook $\mathcal{M}_{distance}$ based on the angle set $\mathbf{\Psi}$. It is noteworthy that the generation process of the distance-adaptive codebook merely involves superimposing pre-designed single-beam patterns from $\mathcal{M}_{singlebeam}$, where the time overhead is negligible. The BS sequentially uses $J$ multi-beam codewords with different distance coverage in $\mathcal{M}_{distance}$ to configure the amplitude of the RHS elements, while the digital beamformer remains constant as $\mathbf{v}_k= \sqrt{\frac{P}{LK}}[1,\cdots,1]^H$~\cite{digital}. After receiving the signals, user $k$ records the received signal power and ultimately feeds back the codeword index $\zeta_{k}$ that maximizes the received power.
Hence, the distance of user $k$ is obtained by
\begin{equation}\label{mu}
\hat{\mu_k}= \mu_{min}+\frac{\mu_{max}-\mu_{min}}{2J}\cdot(2\zeta_{k}-1).
\end{equation}

\begin{figure*}[t]
\setlength{\abovecaptionskip}{0pt}
\setlength{\belowcaptionskip}{0pt}
	\centering
    \includegraphics[width=0.95\textwidth]{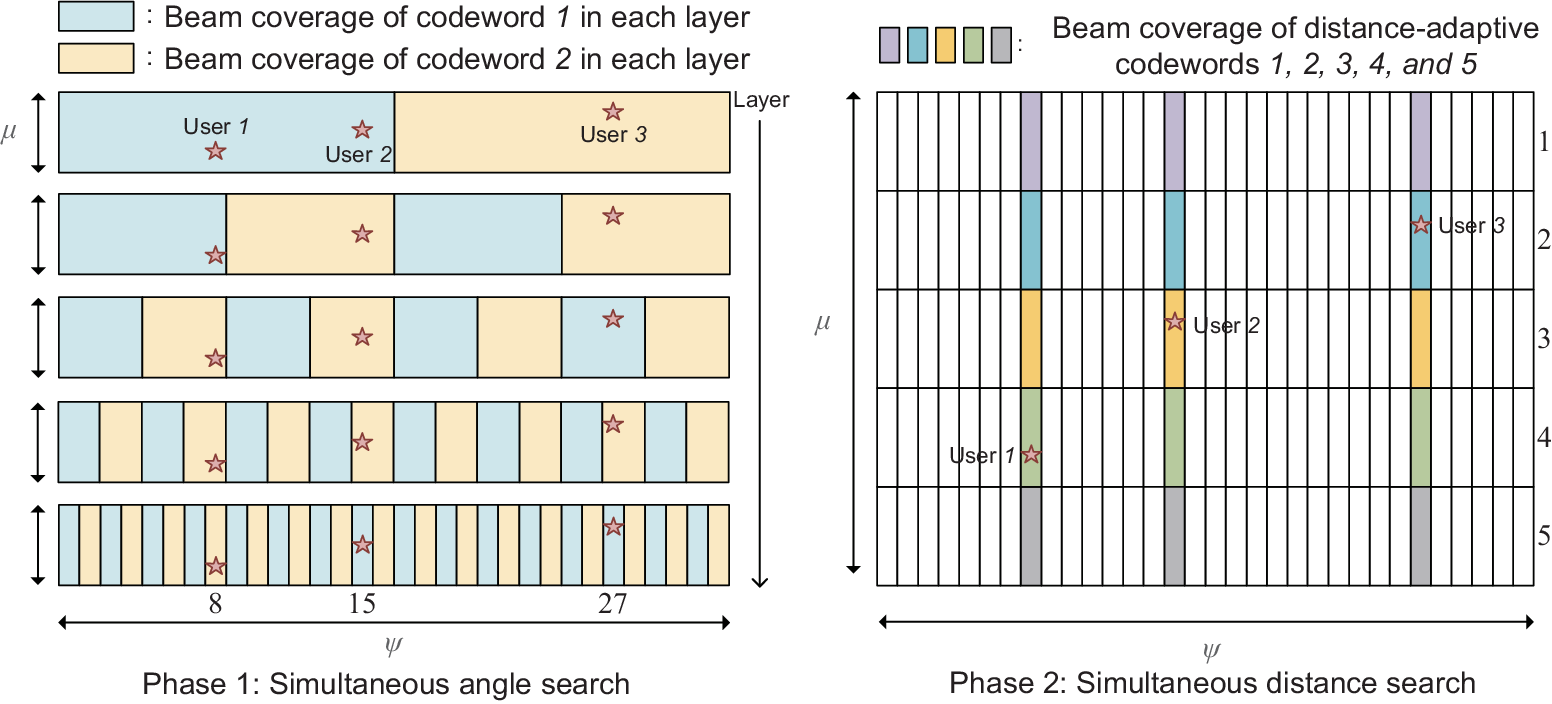}
	\caption{Example: 3-user simultaneous beam training.}
	\label{Fig:example}
\vspace{0em}
\end{figure*}

The overhead of the proposed beam training is $\log_2{I}+J$, where $I$ and $J$ are the numbers of samplings in the angle and distance domain respectively. 
The training overhead of the proposed scheme does not scale with the number of users, enabling low-overhead beam training even in multi-user communication systems.
Table~\ref{overhead} summarizes the training overhead of conventional near-field and far-field beam training schemes.
Notably, the overhead of exhaustive beam training and far-field multi-user beam training also remains constant regardless of the user count.
However, exhaustive search incurs excessively high overheads, while far-field multi-user beam training is inapplicable to near-field users, which will be elaborated in Section \ref{Sec:simulation}.
The table presents the training overhead of various schemes in a 10-user scenario, where the number of angular and distance samplings are set to 64 and 10, respectively.

\renewcommand\arraystretch{1.2} 
\begin{table}[t]
\label{overhead}
\centering
\caption{Beam training overhead of different schemes.}
\resizebox{\columnwidth}{!}{%
\begin{tabular}{|c|c|c|}
\hline
\textbf{Beam training scheme}     & \textbf{Overhead}          & \textbf{Value} \\ \hline
Two-stage beam training\cite{twostage1},\cite{twostage2}           & $\log_2(I)+JK$ &    106    \\ \hline
DFT-distance beam training\cite{dft-dis}        & $I+QJK$             &  364      \\ \hline
Near-field exhaustive search\cite{exh}      &    $IJ$                    &  640     \\ \hline
Far-field multi-user beam training\cite{far_multi-user} &    $\log_2I$                      &  6     \\ \hline
Proposed multi-user beam training  &    $\log_2I+J$                   &  16     \\ \hline
\end{tabular}%
}
\end{table}

\subsection{Illustrative Example: $3$-User Simultaneous Beam Training}

Now we describe the process of the two-phase simultaneous multi-user beam training with an example. Suppose there are three users located at star positions in Fig.~\ref{Fig:example}, each with distinct angles and distances. During the first-phase multi-user beam training, the BS sequentially configures the RHS using two codewords in each layer of the angular codebook $\mathcal{M}_{angle}$. After receiving signals, each user records the codeword index which brings a higher signal power in each layer. Assuming there are $5$ layers in the angular codebook $\mathcal{M}_{angle}$, the bottom layer comprises $32$ fine beams of different angles. After the $5$-layer beam training, User $1$ records codeword indices as $[1,1,2,2,2]$, User $2$ as $[1,2,2,2,1]$, and User $3$ as $[2,2,1,2,1]$.
According to Eq.~\eqref{index}, we can obtain the corresponding angular indices at the bottom layer of the codebook for each user as $[8,15,27]$, which are then fed back to the BS.

After obtaining the angular information of the users, the BS generates the distance-adaptive codebook $\mathcal{M}_{distance}$ corresponding to these three angles in real time.
Assuming there are $5$ samplings for distance, codebook $\mathcal{M}_{distance}$ contains $5$ multi-beam codewords, where each codeword is designed by superimposing $3$ single-beam patterns corresponding to the respective angles and distances from $\mathcal{M}_{singlebeam}$. The BS sequentially configures the RHS with these $5$ codewords, and users record the codeword index that maximizes signal power. Subsequently, the three users feed back the distance codeword indices as $[4,3,2]$.

Finally, the BS obtains user coordinates as $(8,4)$, $(15,3)$, and $(27,2)$, respectively.
Then the angles and distances of users can be calculated according to \eqref{psi} and~\eqref{mu}.

\subsection{Holographic-Pattern Based Beamformer Design}
Through codebook design and beam training, the BS acquires the position of all users and the optimal holographic patterns. To maximize the sum rate, we first design the RHS holographic beamformer $\mathbf{M}$ based on the optimal patterns of all users, and then design the digital beamformer $\mathbf{V}$ to alleviate inter-user interference.

After beam training, the BS acquires the optimal codewords for all users. By superimposing the holographic patterns of each codeword, the RHS holographic beamformer capable of serving all users is generated.
Assuming the coordinates of all users are $[(\hat{\psi_1},\hat{\mu_1}),\cdots,(\hat{\psi_k},\hat{\mu_k})]$, the optimal holographic patterns are represented by $[\mathbf{m}_1^*,\cdots,\mathbf{m}_K^*]$, where each beam pattern is chosen from codebook $\mathcal{M}_{singlebeam}$.
The RHS holographic beamformer can be rewritten as $\mathbf{M}=\text{diag}(\mathbf{m})\mathbf{F}$, where the $(n,l)$ element of $\mathbf{F}\in\mathbb{C}^{N\times L}$ is $\sqrt{\eta}\cdot e^{-\alpha|\mathbf{d}_{n,l}|}\cdot e^{-j\mathbf{k}_s\cdot\mathbf{d}_{n,l}}$. Hence, based on the holographic principle, the RHS holographic beamformer $\mathbf{M}$ is designed as
\begin{equation}
\mathbf{M}=\frac{1}{K}\text{diag}(\sum_{k=1}^K \mathbf{m}_k)\mathbf{F}.
\end{equation}

After determining the RHS beamformer, for the digital beamformer $\mathbf{V}$, we employ the ZF algorithm together with power allocation as an example to alleviate inter-user interference and maximize the sum rate. The ZF precoder serves as a near-optimal solution, characterized by its low complexity.
The digital beamformer $\mathbf{V}$ can be expressed as
\begin{equation}
\mathbf{V}=\mathbf{Q}^H(\mathbf{Q}\mathbf{Q}^H)^{-1}\mathbf{P}^{\frac12}=\widetilde{\mathbf{V}}\mathbf{P}^{\frac12},
\end{equation}
where $\mathbf{Q} = (\mathbf{h}_1\mathbf{M},\cdots,\mathbf{h}_K\mathbf{M}) \in \mathbb{C}^{K\times L}$ and $\mathbf{P} = \mathrm{diag}(p_{1},\cdots,p_{K})$ is the power allocation matrix.
According to the properties of the ZF precoder, the received signal power $|\mathbf{h}_k\mathbf{M}\mathbf{v}_k|^2={p_{k}}$ and the interference $|\mathbf{h}_k\mathbf{M}\mathbf{v}_{k'}|^2=0$.
Therefore, the power allocation problem can be expressed as~\cite{HDMA}
\begin{equation}
\begin{aligned}
&\max_{\mathbf{P}}\quad \sum_{k=1}^{K} \log_{2}{(1+\frac{p_k}{\sigma^2})}\\
&\begin{array}{r@{\quad}r@{}l@{\quad}l}
\mathbf{s.t.}
&  \quad &Tr(\mathbf{V}^H\mathbf{V})= P. \\
\end{array}
\end{aligned}
\end{equation}
The optimal $[p_1,\cdots,p_K]$ are obtained by water-filling~\cite{RIS1}
\begin{equation}
p_k=\frac{1}{\mu_k}\max\{\frac{1}{\xi}-\mu_k\sigma^2,0\},
\end{equation}
where $\mu_k$ is the $k$-th diagonal element of $\widetilde{\mathbf{V}}^H \mathbf{M}^H \mathbf{M} \mathbf{V}$,  and $\xi$ is a normalized factor satisfying $\sum_{k=1}^K\max\{\frac{1}{\xi}-\mu_k\sigma^2,0\}=P$.

\section{Simulation Results}\label{Sec:simulation}

\subsection{Parameter Setting}
In the simulation parameter settings, the carrier frequency is set at $30$ GHz, and the RHS is equipped with $256$ elements, the corresponding Rayleigh distance of which is $81.28$ m. To simultaneously serve both near-field and far-field users, the coverage of the designed codebook is configured to be $\psi \in [-0.5,0.5]$ and $\mu \in [0.005,0.33]$, corresponding to $r \in [3m,150m]$. The number of RF chains is $10$ to support multiple data streams. 
In the codeword design, the normalized desired beamforming gain $D=1$. The SNR is defined as $\frac{1}{\sigma^2}$.
The EM wave propagation factor of RHS $\mathbf{k}_s$ is set as $\frac{2\sqrt{3}\pi}{\lambda}$, and the path loss factor on RHS is assumed as 2.
The iterations in Algorithm~\ref{alg:codeword} and~\ref{alg:single_codeword} is 20 to achive convergence~\cite{HDMA},\cite{holoprin},\cite{imp}.
To evaluate the performance of the proposed multi-user beam training, we compare it with the following schemes.
Assuming the number of samplings in the angle domain and distance domain are $I$ and $J$, respectively, the number of layers in the hierarchical codebook is $S$, and there are $K$ users.
\begin{itemize}
\item[1)] \textbf{Far-field multi-user beam training}~\cite{far_multi-user}: Multi-user simultaneous beam training scheme based on the far-field planar wave model. 
The beam training overhead is $S$.
\item[2)] \textbf{Two-stage Beam Training}~\cite{twostage1},\cite{twostage2}: This scheme first conducts an angle search for users based on a hierarchical codebook, followed by an exhaustive distance search. The beam training overhead is given by $(S+J)\cdot K$.
\item[3)] \textbf{DFT-distance Beam Training}~\cite{dft-dis}: This approach involves the angular search based on a DFT codebook, followed by an exhaustive distance search within candidate angles. The beam training overhead is calculated as $I+Q\cdot J\cdot K$, where $Q$ represents the number of candidate angles.
\item[4)] \textbf{Far-field Exhaustive Beam Training}~\cite{codebook2}: It sequentially searches each angle based on the DFT codebook, whose beam training overhead is $I$.
\item[5)] \textbf{Near-field Exhaustive Beam Training}~\cite{exh}: It sequentially scans each angle and distance, resulting in a beam training overhead of $I\cdot J$.
\end{itemize}

\subsection{Holographic Patterns of Multi-user Angular Codebook}
\begin{figure}[t]
\setlength{\abovecaptionskip}{0pt}
\setlength{\belowcaptionskip}{0pt}
	\centering
    \includegraphics[width=0.48\textwidth]{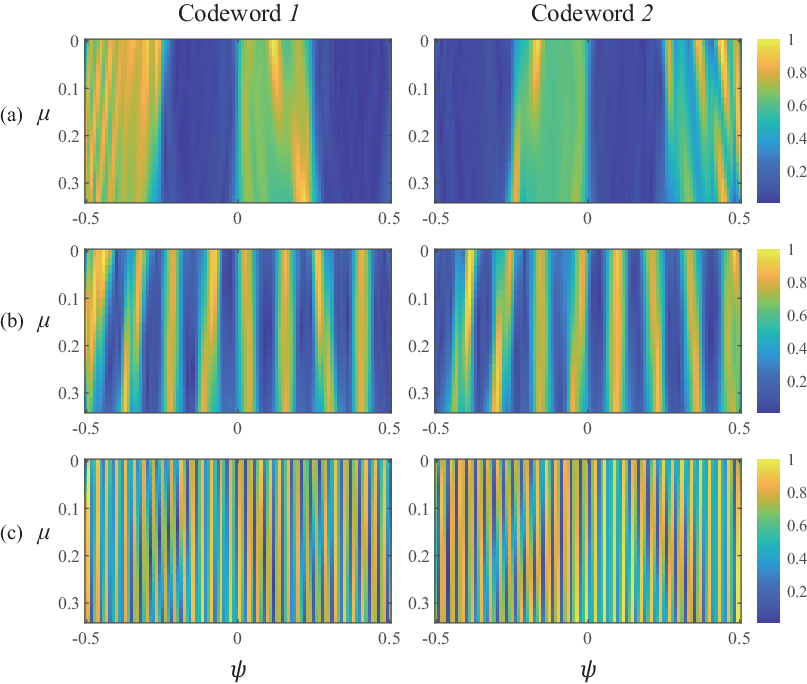}
	\caption{Beam coverage of codeword $1$ and codeword $2$ in (a) the 2-nd layer, (b) the 4-th layer, and (c) the 6-th layer.}
	\label{Fig:codeword_gain}
\vspace{0em}
\end{figure}

The holographic patterns of the codewords in the $2$-nd, $4$-th, and $6$-th layers are illustrated in Fig.~\ref{Fig:codeword_gain}.
The multi-user angular codebook $\mathcal{M}_{angle}$ is designed based on the details in Section~\ref{Sec:angular_book}.
Considering the limitation of the finest angular beams that can be generated by the system, as specified in ~\eqref{bottom}, the number of layers in this codebook is set as $6$.
It can be observed that the beam coverage and beamforming gain of the codewords align well with Fig.\ref{Fig:angularbook} and \ref{Fig:codeword}, respectively, which demonstrates the effectiveness of the proposed codeword design Algorithm~\ref{alg:codeword}.
Furthermore, each codeword exhibits high beamforming gain at different distances within its corresponding angular range, ensuring high received signal power for both near-field and far-field users, thereby enabling simultaneous angular search in the hybrid near-far field.

\subsection{Holographic Patterns of Distance-Adaptive Codebook}
\begin{figure}[t]
\setlength{\abovecaptionskip}{0pt}
\setlength{\belowcaptionskip}{0pt}
	\centering
    \includegraphics[width=0.48\textwidth]{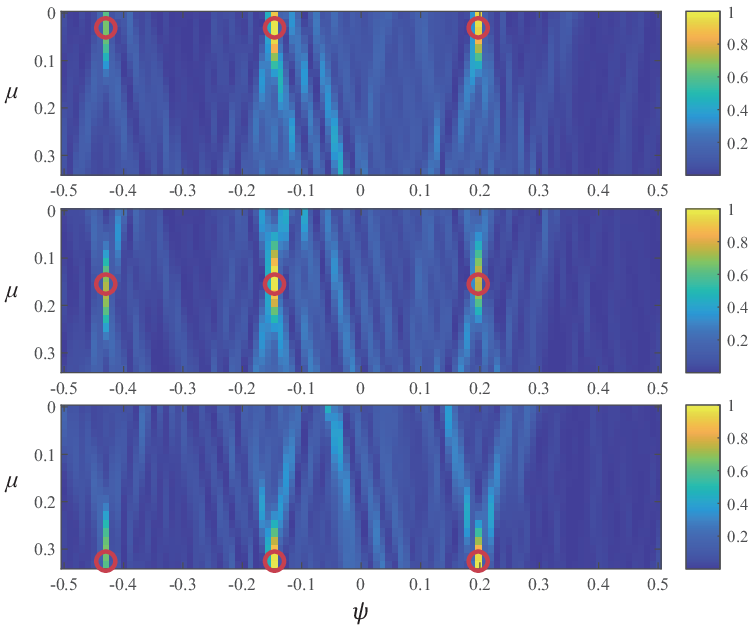}
	\caption{Beam coverage of distance-adaptive (a) codeword $1$, (b) codeword $5$, (c) codeword $10$.}
	\label{Fig:adaptive_codeword}
\vspace{0em}
\end{figure}

Fig.~\ref{Fig:adaptive_codeword} presents the holographic patterns of three codewords with distance indices 1, 5, and 10, respectively.
Through the superposition of holographic patterns, the distance-adaptive codebook corresponding to all users' angles can be expeditiously generated to search the user distance.
Taking simultaneous beam training for three users as an example, suppose the BS obtains beam indices 5, 23, and 45 for the three users at the bottom level of the angular codebook, respectively. In the distance domain, $\mu$ is uniformly sampled at 10 points, and 10 adaptive codewords are generated corresponding to different distances.
It can be observed that each codeword exhibits high gain at a specific distance within these three user angles, validating the effectiveness of the distance-adaptive codeword Algorithm~\ref{alg:distance_codebook} and holographic pattern superposition. By sequentially employing these 10 distance-adaptive codewords, the BS can determine the distance information of all users simultaneously.

\subsection{Comparison with Traditional Far-Field Multi-user Beam Training}

\begin{figure}[t]
\setlength{\abovecaptionskip}{0pt}
\setlength{\belowcaptionskip}{0pt}
	\centering
    \includegraphics[width=0.49\textwidth]{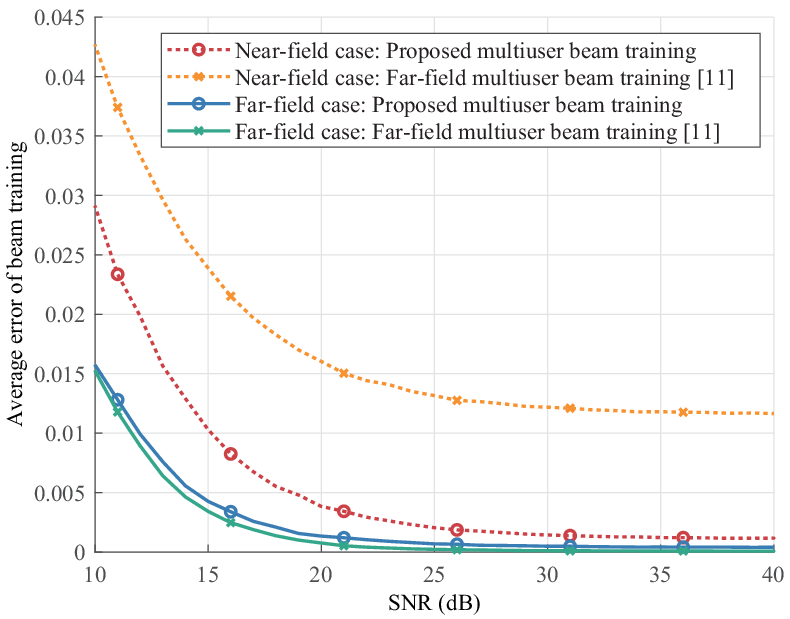}
	\caption{Average error of beam training vs. SNR.}
	\label{Fig:Traditional_Far}
\vspace{0em}
\end{figure}

\begin{figure}[t]
\setlength{\abovecaptionskip}{0pt}
\setlength{\belowcaptionskip}{0pt}
	\centering
    \includegraphics[width=0.49\textwidth]{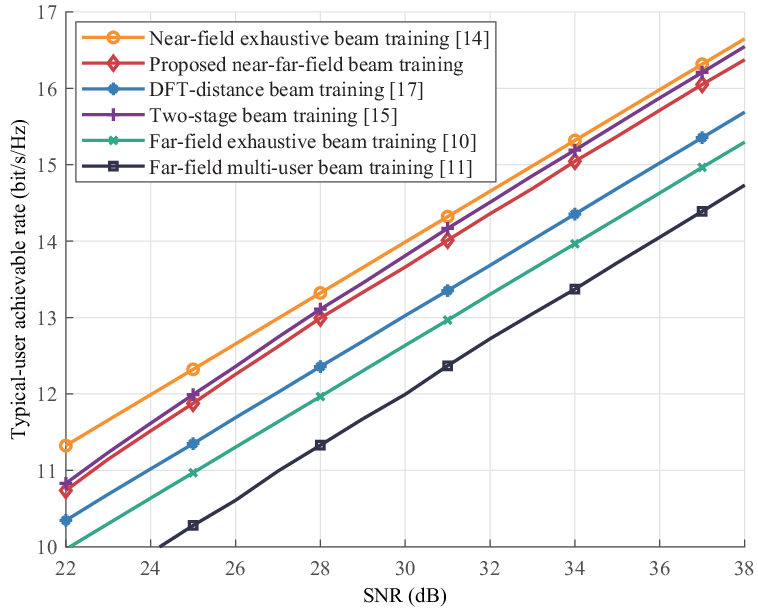}
	\caption{Achievable rate vs. SNR in RHS-enabled systems.}
	\label{Fig:Typical_Rate}
\vspace{0em}
\end{figure}

To evaluate the adaptability of the proposed near-far-field multi-user simultaneous beam training, Fig.\ref{Fig:Traditional_Far} compares with the traditional far-field scheme in the following two typical scenarios: 1) Typical near-field scenario: users are randomly distributed within the near-field region of $r\in[3m, 20m]$. 2) Typical far-field scenario: users are randomly distributed within the far-field region of $r\in[80m, 150m]$.
The beam training error is defined as the distance in the psi-mu domain between the user's actual position $(\psi_k,\mu_k)$ and the beam training result $[(\hat{\psi_k},\hat{\mu_k})$, i.e. $(\psi_k-\hat{\psi_k})^2+(\mu_k-\hat{\mu_k})^2$.
It is observed that in the near-field scenario, the average beam training error of the proposed scheme is significantly lower than that of traditional far-field simultaneous beam training.
This is because the proposed scheme can simultaneously perform both angle and distance search for all users, whereas traditional schemes can only roughly search the user angle. In the far-field scenario, the average beam training error of the proposed scheme is almost identical to that of traditional far-field simultaneous beam training. This is because users are distributed in the far field, far away from the BS, resulting in a large distance $r_k$ (i.e., a small $\mu_k$).
Traditional far-field beam training is enough to obtain user positions, where the $\hat{\mu_k}$ is assumed as 0.
It verifies the effectiveness of the proposed scheme in both near- and far-field scenarios.

\subsection{Comparison with Traditional Beam Training Schemes}
As illustrated in Fig.~\ref{Fig:Typical_Rate}, we compare the typical-user data rate with existing beam training schemes. It can be observed that the proposed multi-user beam training scheme achieves a higher data rate compared to the traditional DFT-distance and far-field approaches. When users are distributed in both the near and far fields simultaneously, these schemes fail to align the channels for near-field users, resulting in a decline in the data rate. Given the same codebook coverage, the data rate of the proposed multi-user scheme is close to that of the two-stage beam training scheme designed for single-user scenarios.
The codeword employed in the two-stage scheme can generate a single-lobe beam that serves only one user, thus resulting in a smaller beam training error and slightly higher data rate.
However, in the multi-user system, the training overhead of the proposed scheme is much lower than that of both the single-user two-stage and exhaustive search methods.

\subsection{Comparison with PA-Enabled Near-Field Beam Training Schemes}

\begin{figure}[t]
\setlength{\abovecaptionskip}{0pt}
\setlength{\belowcaptionskip}{0pt}
	\centering
    \includegraphics[width=0.5\textwidth]{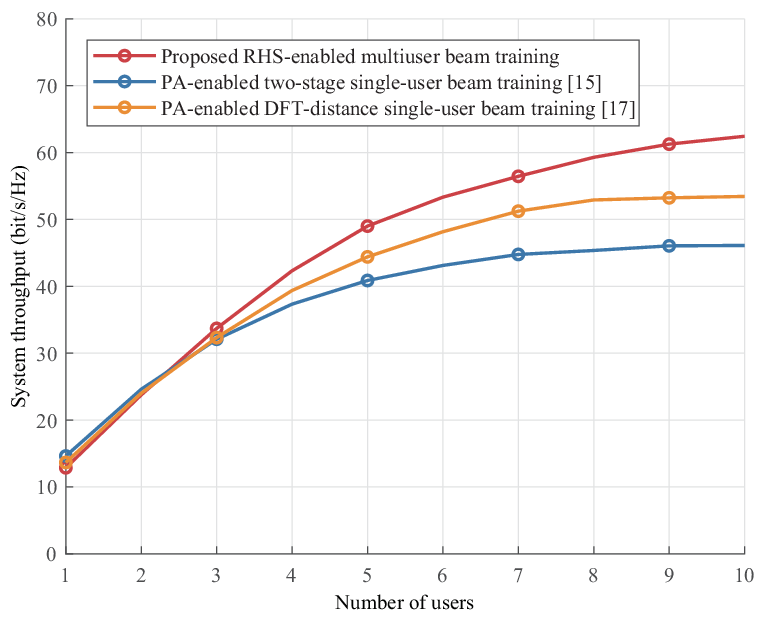}
		\caption{Near-far-field case ($r\in[3m, 150m]$): system throughput vs. the user number.}
	\label{Fig:Traditional_Near1}
\vspace{0em}
\end{figure}

\begin{figure}[t]
\setlength{\abovecaptionskip}{0pt}
\setlength{\belowcaptionskip}{0pt}
	\centering
    \includegraphics[width=0.5\textwidth]{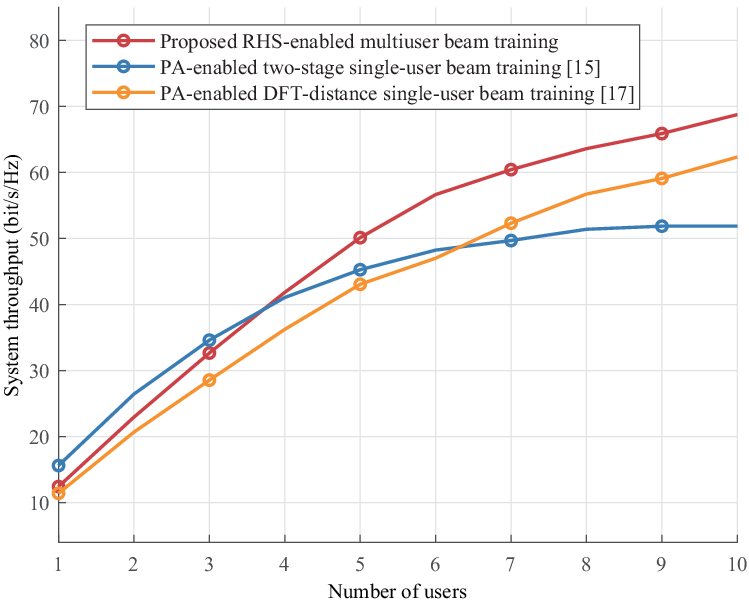}
		\caption{Near-field case ($r\in[3m, 20m]$): system throughput vs. the user number.}
	\label{Fig:Traditional_Near2}
\vspace{0em}
\end{figure}

To evaluate the impact of the proposed low-overhead multi-user beam training scheme on system throughput, we compare its performance with existing near-field beam training approaches enabled by PAs with the same aperture size and input power.
System throughput, which takes into account both beam training overhead and sum data rate, serves as an effective metric to assess system capacity and is defined as $(1-t/T)\times R$, where $t$ represents the time slots of beam training overhead, $T$ the total transmission slots of the system, and $R$ the sum data rate~\cite{slot1}.
We set SNR = 20dB, $T = 0.2 ms$, and a time slot is $0.4 \mu s$\cite{slot2}.

As shown in Fig.~\ref{Fig:Traditional_Near1} and Fig.~\ref{Fig:Traditional_Near2}, we consider two scenarios where users are distributed in the near-field and the near-far-field, respectively.
It can be found that the proposed multi-user beam training scheme achieves higher system throughput as the number of users exceeds 3.
As the number of users increases, the proposed simultaneous multi-user beam training scheme yields higher system throughput with significantly lower beam training overhead. It requires only one-shot beam training for all users, whereas the training overhead of traditional schemes continuously increases with the number of users.
Since the overhead of the simultaneous multi-user training scheme does not scale with the number of users, the proposed scheme can improve the system throughput of multi-user communication systems aided by large-scale arrays.

\begin{figure}[t]
\setlength{\abovecaptionskip}{0pt}
\setlength{\belowcaptionskip}{0pt}
	\centering
    \includegraphics[width=0.49\textwidth]{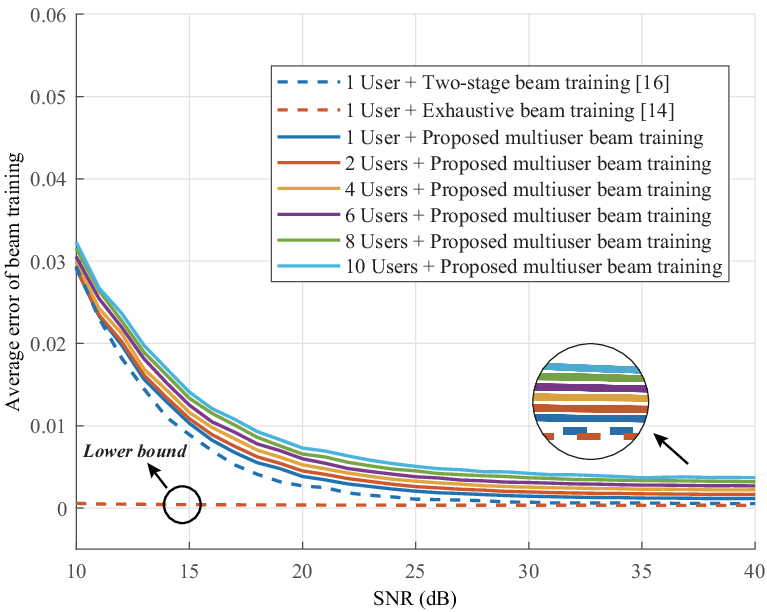}
	\caption{Average error of beam training vs. SNR given different numbers of users.}
	\label{Fig:User_number}
\vspace{0em}
\end{figure}

\subsection{Effect of Number of Users on Multi-User Beam Training}
Now we focus on the impact of user count on beam training errors in near-field scenarios. Fig.~\ref{Fig:User_number} illustrates the average beam training error for different numbers of users and beam training schemes.
The first phase of the proposed multi-user beam training, i.e. the multi-user angle search, is unaffected by the number of users, as it utilizes a pre-designed codebook whose codewords do not vary with the user count.
The number of users has an impact on the second phase of beam training, i.e., multi-user distance search.
It can be observed that as the number of users increases, the average error of the proposed beam training scheme slightly rises.
This is attributed to the fact that when generating distance-adaptive codewords, an increase in user count necessitates covering a wider range of angles, which reduces the beam gain at each angle givin the power constraint.
With the increase in SNR, the multi-user beam training scheme can approach the lower bound of exhaustive search with significantly lower training overhead.

\section{Conclusions}\label{Sec:conclusions}
In this paper, we propose an RHS-aided multi-user beam training scheme for the hybrid near-far field. The beam training process consists of two phases: simultaneous angle search and distance search, where all users are concurrently served in the one-shot beam training.
Based on the holographic principle, the multi-user angular codebook and distance-adaptive codebook are presented. Unlike conventional far-field multi-user angular beam training, the proposed codebook can simultaneously support both near-field and far-field users.
The overhead of the proposed multi-user beam training scheme does not scale with the number of users, significantly lower than that of traditional near-field single-user beam training approaches.

Simulation results demonstrate that: 1) The codewords in the multi-user angular codebook can cover all distances at a specific angle, and each distance-adaptive codeword can cover a specific distance across all user angles, validating the effectiveness of the proposed codeword generation method. 2) As the SNR increases, the accuracy of the proposed near-far-field multi-user beam training approaches the upper bound of exhaustive search with significantly reduced overhead, and outperforms the conventional far-field multi-user beam training scheme. 3) The overhead of the proposed multi-user simultaneous beam training scheme remains constant regardless of the number of users, thereby improving the system throughput compared to traditional near-field beam training schemes.

\end{document}